\documentclass[12pt]{article}
\usepackage{amsmath}
\usepackage{amsfonts}
\begin{document}
\def\eg{{\em e.g.\,}}
\def\ie{{\em i.e.\,}}
\def\C{\mathrm{C}}
\def\Q{\mathrm{Q}}
\def\cH{\mathcal{H}}
\def\dof{{d.o.f.\,}}
\def\bra{\langle}
\def\ket{\rangle}
\def\av#1{\langle #1 \rangle}
\def\vi{\mathbf{i}}
\def\vk{\mathbf{k}}
\title{On some algebraic problems arising in quantum mechanical 
description of biological systems}
\author{M.V.Altaisky\thanks{Joint Institute for Nuclear Research,
 Dubna, 141980, Russia;
 and Space Research Institute RAS, Profsoyuznaya 84/32, 
Moscow, 117997, Russia, altaisky@mx.iki.rssi.ru.}}
\date{}
\maketitle
\begin{abstract}
The biological hierarchy and the differences between living and non-living 
systems are considered from the standpoint of quantum mechanics. The 
hierarchical organization of biological systems requires hierarchical 
organization of quantum states. The construction of 
the hierarchical space of state vectors is presented. The application 
of similar structures to quantum information processing is considered.
\end{abstract}

\section{Quantum Mechanics and Evolution}
The most important discoverings in natural sciences are connected to 
quantum mechanics. There is also a bias that 
biological phenomena will be explained by quantum theory in future, 
since quantum theory contains all basic principles of particle 
interactions and these principle had success in molecular dynamics, the 
basis of life. 
Biology has however a number of concepts and facts not displayed explicitly 
in inanimate world. The following 
are of principle importance:
\begin{enumerate}
\item   The properties of a living system are more than just a collection 
        of its components properties. In other words, it is impossible to 
        predict the whole set of properties of a complex biological system 
        even having known all properties of its components.
\item   The properties and functions of the components of a system depend 
        on the state of the whole system. In other words, the same components 
        being included in different systems may have different properties.
\item   There is an {\sl Evolution} --- a process of creating new entities,
        forms and functions on the base of the existing components. 
\end{enumerate} 
For both biology and quantum physics the relation 
{\sf ``the part - the whole''} is extremely important, 
and not trivial: the factorization  
$ \Psi(x) = \prod_i \psi_i(x_i) $ holds only for the systems of 
noninteracting particles. 

The properties (2,3) listed above are implicitly 
based on the {\sf concept of scale}: an entity to become a part of another 
entity should be in some metric smaller than it. 
If the metric is Euclidean, or at least Archimedian, the evolution of the 
Universe can be said to go from small scales to large scales. 
In this sense, the elementary particles and atoms 
had had their evolution: at early times of the Universe  the 
nucleons had been built of quarks, the nuclei from nucleons and so long. 

We do not have an answer to the question, 
why the evolution had taken the way it has been going through.
However, {\em if the whole is more than the sum of parts and the properties 
of the parts depend on the state of the whole, there are some implications 
for quantum mechanics}.

To describe a state of an object 
$A_1$ (interacting with objects $A_2,\ldots,A_N$), which is a part of 
an object $B_1$ we have to write the wave function in the form 
$
\{ \Psi_{B_1},  \Psi_{B_1A_1} \},
$
where $\Psi_{B_1}$ is the wave function of the whole, 
and $\Psi_{B_1A_1}$ is the wave function of a {\sl 
component} $A_1$ belonging to the entity $B_1$.
For instance, $A_1,A_2,A_3$ may be quarks, and $B_1$ may be proton.
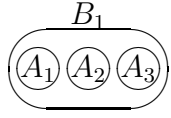
\begin{figure}
\begin{center}
\begin{picture}(80,50)
\put(40,40){\makebox(0,0){$B_1$}}
\put(40,20){\oval(60,30)}
\put(21,20){\circle{17}}
\put(40,20){\circle{17}}
\put(59,20){\circle{17}}
\put(21,20){\makebox(0,0){$A_1$}}
\put(40,20){\makebox(0,0){$A_2$}}
\put(59,20){\makebox(0,0){$A_3$}}
\end{picture}
\end{center}
\caption{The whole $B_1$ and its parts $A_1,A_2,A_3$.}
\end{figure} 
The objects $A_1,\ldots,A_N$ are {\em inside} $B_1$ and it is 
impossible to commute $[\Psi_{B_1},\Psi_{B_1A_1}]$ or 
to multiply them $\Psi_{B_1},\Psi_{B_1A_1}$, since {\em the functions 
$\Psi_{B_1}(x)$ and  $\Psi_{B_1A_1}(x)$ live in different functional 
spaces}. 

We generally observe,  
that {\sf each hierarchical level has its own symmetry}. This is $SU_3$ 
for quark level or isospin group for nuclei etc. So, each hierarchy level 
can be described by   
$(G_I,X^{G_I}),$  
where $I$ labels the scale, $G_I$ is the symmetry group at 
this scale, $X^{G_I}$ is a topology on $G_I$.  
The Euclidean space is a particular case of the translation group 
$G_I : x \to x +b.$
The wave function of an object $B^{I_1}$ of the level $I_1$ consisting 
of $N$ objects $\{A_i^{I_2}\}_{i=1,N}$ can be written as 
\begin{equation}
\Psi_B =  \Bigl\{ 
        \psi^{I_1}_B(x^{G_{I_1}}),
        \{\psi^{I_2}_{BA_1}(x^{G_{I_2}}_1),\ldots,
        \psi^{I_2}_{BA_N}(x^{G_{I_2}}_N) \},\ldots
        \Bigr\}.
\label{hier}
\end{equation}

In biology, as well as in physics, the symmetry breaking plays an 
important role. It is known, that the amount of information written 
in DNA, if calculated as one nucleotide -- one bit, is  
insufficient to describe  the formation of adult organism. Thus, 
the information is likely to be written more effectively than just 
a technical plan of the organism. What is encoded, is probably a chain 
of bifurcation points to be undergone in growth process. 

If the quantum mechanics is valid on the macroscopic scales, we can say that 
the hierarchy levels emerge as a result 
of a) evolution, b) self-organization, and 
formation of new entities from existing ones 
$\{ \emptyset; \psi_{A_1} \otimes \psi_{A_2}\} \to 
   \{ \psi_{B}; \{ \psi_{BA_1},\psi_{BA_2}\}\}$,
where the empty-set $\emptyset$ denotes the 
non-existing common ``container'' for two 
components $\psi_{A_1}$ and $\psi_{A_2}$; $\psi_{B}$ is a new entity 
formed by $\psi_{A_1}$ and $\psi_{A_2}$. 

The hierarchy of biological system joints the hierarchy of non-living matter 
by means of the {\sl cell} -- unit of life, and its part, the genome, 
the sequence  of macromolecules which prescribes  the evolution of all living systems, from
cell to organism. \vskip0.1cm
\begin{center}
{\sf The hierarchy levels of living and non-living matter}
\vskip0.1cm
\begin{tabular}{|r|r|}
\hline
Living matter & Non-living matter \\
\hline 
\dots     & \\
ecosystem &            \\
population&  \\
organism  & \\
organ     & \\
\multicolumn{2}{|c|}{cell}       \\
\multicolumn{2}{|c|}{organell}   \\
\multicolumn{2}{|c|}{genome} \\
          & molecule \\
          & atom \\
          & nucleus \\
          & nucleon \\
          & \dots  \\
\hline
\end{tabular}
\end{center}  
The place on an entity in the hierarchy tree and its distance from the 
position of the organism it belongs to determines the dynamical 
repertoire of the entity. 

The evolutional distance between maximal and 
minimal parts of the organism determines its ability of self-recovering. 
For instance, if one end of {\em Hydra oligactis} (a simplest animal 
living in water) 
is cutted off, the remaining cells react to the absence of this part by 
rearranging themselves and giving growth to new cells and form a complete 
animal. This process involves at least three levels:
 
\centerline{Organism $\longrightarrow$ Cell $\longrightarrow$ Cell component.}    
Let $T(G_A)$ be an irreducable representation of the symmetry group $G_A$ 
which corresponds to the wave function of the intact organism. 
The self-repairation process can be then described 
by following 3 level diagram:
\begin{eqnarray}
\nonumber \{ \psi_A; \{\psi_{AC_1}, \ldots , \psi_{AC_N} \}\} + \Gamma \to
\{ \psi_A; \{\psi_{AC_1}, \ldots , \psi_{AC_K} \}\} +
\{\psi_{C_{K+1}}, \ldots , \psi_{C_N} \} \\ 
\to \{ \psi_A; \{\psi_{AC_1c_1},\psi_{AC_1c_2} \ldots , \psi_{AC_Kc_L} \}\}
\label{restr} \\ 
\nonumber \to \{ \psi_A; \{\psi_{AC_1}, \ldots , \psi_{AC_N} \}\} 
\end{eqnarray} 
On the first stage, affected by destructive action $\Gamma$, 
the part of the system, a block of $(N-K)$ cells is cutted of. The 
{\sl remainder} 
\begin{equation}
\{ \psi_A; \{\psi_{AC_1}, \ldots , \psi_{AC_K} \}\}
\label{remainder}
\end{equation}
 does not form a complete organism any longer; the product of representations 
$\displaystyle \bigotimes_{i=1}^K T(G_{C_i})$ 
does not contain  $T(G_{A})$. 
So, the wave function of the remainder (\ref{remainder}) breaks down 
to the third level hierarchy wave function  
$\{ \psi_A; \{\psi_{AC_1c_1},\psi_{AC_1c_2} \ldots , \psi_{AC_Kc_L} \}\},$
providing a possibility of building a  representation tensor product, 
which contains $T(G_{A})$. On the third level the wave functions are 
being rearanged according to this tensor product and the missed second 
level blocks are being rebuilt.

The observations at all levels of the evolutional hierarchy, from simple 
organnels to complex ecosystems suggest that basically,  {\sf only neighboring 
levels interact}. The quantum nature of the interactions 
in this hierarchy may be used to understand, why a cutted skin recovers, but 
the arm cutted of is not being recovered. If the lost part of the organism 
has $M\gg1$ hierarchical levels a $M$-level cascade process should run 
in the remainder to rebuild the lost part. A process of this type will 
require  significant flux of energy, and at the same time  
a tremendous flux of negative entropy, to restore the symmetry
of the wave function of the whole organism by rearanging the wave 
functions of its components.

\section{Pauli principle}
Pauli principle: {\em Two 
fermions of the same system can not be in the same quantum state}. 
This is a consequence of the fact that the wave function of 
a fermion system should be antisymmetric with respect to particle 
transposition.

If we gerard the interaction of neighboring levels only, 
we can say that, {\sf two fermions belonging to the same system  of the 
next hierarchical level can not be in the same state}. Therefore, it is 
impossible for two electrons in atom to have the same quantum numbers 
($n,l,m$), but is it possible for two electrons 
of the same {\em molecule} to be in the same state?
It seems evident that  
two electrons of different macroscopic objects can be in the same state.
{\em But is it really possible for two electrons of the same molecule?}

So, the real question is: what should we really mean by ``the hierarchy 
level next to atom''?
There is no common sense answer to this question, but if the generalization 
of Pauli principle formulated above, is valid, and the only question is 
what is the {\em next hierarchical level}, the matter can be experimentally 
investigated, at least in principle. To some extent, the idea of 
possible experiment of this type have been suggested by D.~Home 
and R.~Chattopadhyaya \cite{HC1996}. 
If biological macromolecules, the DNA, can be 
used as a device for quantum measurement, it  
means that if a photon absorbed by a DNA molecule, the wave function of 
the whole 
molecule flops from one quantum state into another. But the DNA molecule 
itself consists of smaller molecules. So, there are two alternatives: 
either the absorption of a photon changes the wave function of the DNA only 
by changing the wave function of one of its components, or it changes 
the wave function of the whole DNA. In the latter case, due to the 
interaction between the whole and its parts, the absorption of the photon 
at one edge of DNA can can be immediately detected at the opposite edge, 
at least in principle.

\section{Wave functions of living and non-living systems}
The living and non-living systems are different in the {\sf complexity},
in the Kolmogorov sense. The Hamiltonian {\sl for a non-living system}
can be constructed using the representations of the symmetry groups of 
its components and their interactions. This description is shorter than 
a time series of its matrix elements $E_{mn}(t)$ taken at each moment of 
time. {\sl For a living system} the shortest description of the evolution 
operator may be the time seria  $E_{mn}(t)$ itself, or its subseria. 

Let us formulate the difference in the language of group theory:
\begin{enumerate} 
\item We can assert, that {\bf for nonliving systems}  
the knowledge of irreducible representations of the component wave function 
and a symmetry group which accounts for their interaction completely 
determines the wave function of the whole. 
\item The 
wave function of {\bf a living system} is constrained, but not completely 
determined by the representations of symmetry group of its components. 
This means, that even if we know the wave functions of all components 
of a living system we still can not predict the behavior of the 
system without a 
separate knowledge of the next level wave function, i.~e. the wave function of 
the whole. 
\end{enumerate}

To conclude with, we should mention that possible distinction between living 
and non-living systems, itemized above in this paper, makes a new point in 
the Schr\"odinger cat problem 
$$\frac{1}{\sqrt2}|cat\ dead\rangle + \frac{1}{\sqrt2}|cat\ alive\rangle = ?$$
In hierarchical formalism, the wave function of a {\em dead cat} is 
constructed from the direct products of the irreducible representations 
of its parts. The wave function of the {\em alive cat} comprise the 
wave function of the whole cat as well. So these two wave function live in different 
functional spaces. 
\section{Density matrix}
The standard approach to introduce the density 
matrix \cite{Feynman} is to consider the quantum states of the 
``system+environment'': $
|\psi\ket = \sum_{ij} C_{ij}|\phi_i\ket |\theta_j\ket,
$
where $\bigl\{|\phi_i\ket\bigr\}_i$ are the 
state vectors of the system, with $\bigl\{|\theta_i\ket\bigr\}_i$ 
are those of the environment, \ie the rest of the Universe. 
In coordinate representation 
\begin{equation}
\psi(x,y) = \bra y | \bra x | \psi\ket = 
\sum_{ij} C_{ij} \phi_i(x) \theta_j(y) \equiv \sum_i c_i(y)\phi_i(x).
\end{equation}
In any measurement performed on quantum system the wave 
function can be considered as a superposition of the different 
states of the system taken with weights $c_i(y)$ dependent on the state of 
the environment. This means the operators of physical 
observables related to the system act only on $|\phi\ket$ vectors:
$A |\phi_i\ket |\theta_j\ket = \bigl(|\phi_i\ket \bigr)|\theta_j\ket$
and the average value of the observable $A$ is given by 
\begin{equation}
\av{A} \equiv \bra\psi|A|\psi\ket 
       = \sum_{ii'}\rho_{i'i} \bra\phi_i|A|\phi_{i'}\ket \equiv Sp(\rho A) 
\end{equation}
where 
$\rho_{i'i} \equiv \sum_j C_{ij}^* C_{i'j}$
is the density matrix. (The orthonormality of the state vectors of 
the environment is assumed $\bra \theta_i | \theta_j\ket = \delta_{ij}$)

The density matrix $\rho_{i'i}$ being hermitian is usually represented in 
the diagonal form 
$$\rho_{ii'} = \bra \phi_{i'}|\hat\rho|\phi_i \ket, 
\quad \hbox{where\ }  \hat\rho = \sum_i |i\ket\omega_i\bra i|,
\quad Sp\,\rho=\sum_i\omega_i=1,
$$
where the eigenvalue $\omega_i$ is the probability of finding the 
system in the $i$-th eigenstate.

In context of biological systems, we can expect the probability of 
quantum states of the subsystem 
(A) to be dependent basically on the state of the system (B) it is  
embraced in. 
For instance, the state of the nuclei is basically dependent on the state 
of the cell, and much less on the states of other systems; the state of the 
cell in its turn depends on the state of the organ etc. 
There is a temptation to identify $|\theta_j\ket$ with the states 
of the system and $|\phi_i\ket$ with the states of the subsystem. 
This is not the same as in standard quantum mechanics, because 
$
A\cap B = A, A\cap E =\emptyset, A\cup B=B, A\cup E=U
,$
where $A$ is subsystem of $B$, $E$ is the environment of $A$ and $U$ is the 
Universe. 

If the system $B$ consists of $k$ parts $A_1,\ldots A_k$, than we 
can write the wave function of the form 
\begin{equation}
|\psi\ket = \sum C^j_{i_1,\ldots,i_k}|\phi_{i_1}\ket\otimes\ldots\otimes
|\phi_{i_k}\ket |\theta_j\ket
\label{s_i}
\end{equation}
The equation \eqref{s_i} is an approximation accounting all effects of the 
environment 
on the subsystem $A_k$ only by means of the 
effect of the system $B$ to its subsystem $A_k$. We shall call $\phi$ the 
{\em microlevel} an $\theta$ the {\em macrolevel} wave functions.  

Let $A$ be an observable acting on the microlevel of a system containing 
$k$ subparts. Then, as a result of the summation over all macroloevel 
states, the average value of the observable $A$ is 
\begin{equation}
\av{A} = \sum {C^*}^{j'}_{{i_1}',\ldots,{i_k}'}{C}^{j}_{{i_1},\ldots,{i_k}}
\bra\theta_{j'}| \bra \phi_{{i_1}'}|\ldots\bra\phi_{{i_k}'}|
A 
|\phi_{{i_1}}\ket,\ldots,|\phi_{{i_k}}\ket|\theta_j\ket   
      =\rho_{\vi,\vi'}\bra \vi |A|\vi'\ket,
\end{equation}
where $\vi \equiv (i_1,\ldots,i_k), 
|\vi\ket \equiv |\phi_{{i_1}}\ket,\ldots,|\phi_{{i_k}}\ket$ is the multiindex 
of the microlevel state.
If the operator $A=A_1$ acts only on the first ($i_1$) subsystem of 
the microlevel, the density matrix for this subsystem is obtained by the 
averaging over all other ($i_2,\ldots,i_k$) subsystem and the macrosystem 
state
$
\rho^{(1)}_{i_1 {i_1}'}= \sum_{j,i_2,\ldots,i_k} 
{C^*}^j_{{i_1}',i_2,\ldots,i_k}{C}^j_{{i_1},i_2,\ldots,i_k}.
$

In analogy with quantum computing algorithms \cite{qq}, we can introduce  
operators which acts on the microlevel depending on the state of the 
macrolevel 
\begin{equation}
\hat B = |\theta_m\ket B^m_{\vi\vk}\bra\theta_m| |\vi\ket\bra\vk|
\end{equation}
The mean value of the corresponding observable in a two level hierarchical 
system is 
$$
\av{B} = \bra\psi|\hat B|\psi\ket=\sum_{j,\vi,\vi'} {C^*}^j_\vi 
B^j_{\vi\vi'} C^j_{\vi'}.
$$
\section{Recording information in hierarchical quantum systems}
\vskip0.1cm
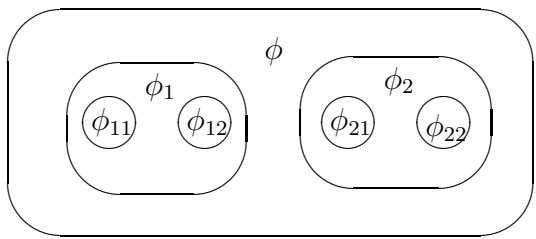
\begin{figure}[ht]
\begin{center}
\setlength{\unitlength}{1974sp}
\begingroup\makeatletter\ifx\SetFigFont\undefined%
\gdef\SetFigFont#1#2#3#4#5{%
  \reset@font\fontsize{#1}{#2pt}%
  \fontfamily{#3}\fontseries{#4}\fontshape{#5}%
  \selectfont}%
\fi\endgroup%
\begin{picture}(6616,2864)(518,-2393)
\thinlines
\put(1801,-961){\circle{618}}
\put(3001,-961){\circle{600}}
\put(4801,-961){\circle{618}}
\put(6001,-961){\circle{600}}
\put(2401,-1036){\oval(2250,1650)}
\put(5401,-961){\oval(2400,1650)}
\put(3826,-961){\oval(6600,2850)}
\put(1576,-961){\makebox(0,0)[lc]{$\phi_{11}$}}
\put(2776,-961){\makebox(0,0)[lc]{$\phi_{12}$}}
\put(4576,-961){\makebox(0,0)[lc]{$\phi_{21}$}}
\put(5776,-1036){\makebox(0,0)[lc]{$\phi_{22}$}}
\put(2251,-511){\makebox(0,0)[lc]{$\phi_1$}}
\put(5251,-436){\makebox(0,0)[lc]{$\phi_2$}}
\put(3751,-61){\makebox(0,0)[lc]{$\phi$}}
\end{picture}
\end{center}
\caption{Hierarchical quantum system used for information recording. 
The state vector of the whole system is 
$
|\Phi\ket =  \Bigl\{ 
        |\phi^{0}\ket,
        \{ |\phi^{1}_{1}\ket,|\phi^{1}_{2}\ket \},
	\{ |\phi^{2}_{11}\ket,|\phi^{2}_{12}\ket,
           |\phi^{2}_{21}\ket,|\phi^{2}_{22}\ket \},
        \Bigr\}. 
$
}
\label{hier-inf:pic}
\end{figure}
The tree-like hirarchical structures can work similar to wavelet based 
data compression.  For example, let us consider 
a set of $l=2^{N-1}$ quantum bits. If these qubits 
are embraced in $N=3$ level hierarchical system, shown in 
Fig.~\ref{hier-inf:pic}, 
then, instead of 4 original qubits 
$\phi^{2}_{11},\phi^{2}_{12},\phi^{2}_{21},\phi^{2}_{22}$, 
on each level of the hierarchy we construct a Haar wavelet \cite{daub} 
like basis:  
\begin{equation}
\begin{array}{ll}
|\phi^1_1\ket = \frac{|\phi^{2}_{11}\ket +|\phi^{2}_{12}\ket }{\sqrt2} &
|\psi^1_1\ket = \frac{|\phi^{2}_{11}\ket -|\phi^{2}_{12}\ket }{\sqrt2} \\
|\phi^1_2\ket = \frac{|\phi^{2}_{21}\ket +|\phi^{2}_{22}\ket }{\sqrt2} &
|\psi^1_2\ket = \frac{|\phi^{2}_{21}\ket -|\phi^{2}_{22}\ket }{\sqrt2} \\
|\phi^0  \ket = \frac{|\phi^{1}_{1} \ket +|\phi^{1}_{2} \ket }{\sqrt2} &
|\psi^0  \ket = \frac{|\phi^{1}_{1} \ket -|\phi^{1}_{2} \ket }{\sqrt2} \\
\end{array}.
\end{equation}
Since the $\phi$ states are linear combinations of $\phi$ and $\psi$ states 
of the previous level, 
finally the whole information is stored in 4 independent states 
of the 2 top levels:
$$
\Psi =  \Bigl\{ 
        |\phi^{0}\ket,|\psi^{0}\ket,
        \{ |\psi^{1}_{1}\ket,|\psi^{1}_{2}\ket \}
        \Bigr\}.
$$


\centerline{***}
The author is thankful to Drs. F.Gareev, T.Gannon and 
O.Mornev for critical reading of the manuscript and useful comments.
This work was supported in part by ISTC project  1813p/2001.

\end{document}